\documentclass[
	aps, rmp,
	superscriptaddress,
	nofootinbib,
	twocolumn,
]{revtex4-1}

\usepackage{amsmath}
\usepackage{amssymb}
\usepackage{graphicx}
\usepackage{units}
\usepackage{hyperref}

\newcommand{\Nl}{{N_{\rm L}}}
\newcommand{\Nr}{{N_{\rm R}}}

\newcommand{\ctot}{c_{\rm tot}}
\newcommand{\ct}{c_{\rm t}}
\newcommand{\cb}{c_{\rm b}}
\newcommand{\sB}{s_{\rm B}}
\newcommand{\xitot}{\xi_{\rm tot}}

\makeatletter{}%
\usepackage{amsmath}
\usepackage{xcolor}
\usepackage{dsfont} %
\usepackage{xspace}

\newcommand{\ie}{\mbox{i.\hspace{0.125em}e.}\@\xspace}
\newcommand{\eg}{\mbox{e.\hspace{0.125em}g.}\@\xspace}

\definecolor{plot1}{RGB}{6,115,183}
\definecolor{plot2}{RGB}{255,118,0}
\definecolor{plot3}{RGB}{0,169,25}
\definecolor{plot4}{RGB}{230,0,28}
\definecolor{plot5}{RGB}{0,0,0}
\definecolor{plotBlue}{RGB}{6,115,183}
\definecolor{plotOrange}{RGB}{255,118,0}
\definecolor{plotGreen}{RGB}{0,169,25}
\definecolor{plotRed}{RGB}{230,0,28}

\newcommand{\Eqref}[1]{\mbox{Eq.\hspace{0.25em}\eqref{#1}}}
\newcommand{\Eqsref}[1]{\mbox{Eqs.\hspace{0.25em}\eqref{#1}}}
\newcommand{\figref}[1]{\mbox{Fig.\hspace{0.25em}\ref{#1}}}

\newcommand{\diff}{\text{d}}

\DeclareMathOperator{\var}{var}

\DeclareMathOperator{\erfc}{erfc}
\newcommand{\mean}[1]{\langle #1 \rangle}

\DeclareFontFamily{U}{mathx}{\hyphenchar\font45}
\DeclareFontShape{U}{mathx}{m}{n}{<-> mathx10}{}
\DeclareSymbolFont{mathx}{U}{mathx}{m}{n}
\DeclareMathAccent{\widebar}{0}{mathx}{"73}
\newcommand{\vect}{\boldsymbol}

\def\barroman#1{\sbox0{#1}\dimen0=\dimexpr\wd0+1pt\relax
  \makebox[\dimen0]{\rlap{\vrule width\dimen0 height 0.06ex depth 0.06ex}%
    \rlap{\vrule width\dimen0 height\dimexpr\ht0+0.03ex\relax 
            depth\dimexpr-\ht0+0.09ex\relax}%
    \kern.5pt#1\kern.5pt}}

\renewcommand{\eqref}[1]{\ref{#1}}

\renewcommand{\figref}[1]{Fig.\nolinebreak[4]\hspace{0.25em}\nolinebreak[4]\ref{#1}}

\newcommand*\patchAmsMathEnvironmentForLineno[1]{%
  \expandafter\let\csname old#1\expandafter\endcsname\csname #1\endcsname
  \expandafter\let\csname oldend#1\expandafter\endcsname\csname end#1\endcsname
  \renewenvironment{#1}%
     {\linenomath\csname old#1\endcsname}%
     {\csname oldend#1\endcsname\endlinenomath}}%
\newcommand*\patchBothAmsMathEnvironmentsForLineno[1]{%
  \patchAmsMathEnvironmentForLineno{#1}%
  \patchAmsMathEnvironmentForLineno{#1*}}%
\makeatletter
\@ifclasswith{revtex4-1}{linenumbers}{
\patchBothAmsMathEnvironmentsForLineno{equation}%
\patchBothAmsMathEnvironmentsForLineno{align}%
\patchBothAmsMathEnvironmentsForLineno{flalign}%
\patchBothAmsMathEnvironmentsForLineno{alignat}%
\patchBothAmsMathEnvironmentsForLineno{gather}%
\patchBothAmsMathEnvironmentsForLineno{multline}%
}
\makeatother

\makeatletter
\let\cat@comma@active\@empty
\makeatother

\begin{document}

\title{Olfactory coding with global inhibition}
\title{Normalized neural representations of natural odors} 

\date{\today}

\author{David Zwicker}
\homepage{http://www.david-zwicker.de}
\affiliation{School of Engineering and Applied Sciences, Harvard University, Cambridge, MA 02138, USA}
\affiliation{Kavli Institute for Bionano Science and Technology, Harvard University, Cambridge, MA 02138, USA}

\begin{abstract}

The olfactory system removes correlations in natural odors using a network of inhibitory neurons in the olfactory bulb.
It has been proposed that this network integrates the response from all olfactory receptors and inhibits them equally.
However, how such global inhibition influences the neural representations of odors is unclear.
Here, we study a simple statistical model of this situation, which leads to concentration-invariant, sparse representations of the odor composition.
We show that the inhibition strength can be tuned to obtain sparse representations that are still useful to discriminate odors that vary in relative concentration, size, and composition.
The model reveals two generic consequences of global inhibition: (i) odors with many molecular species are more difficult to discriminate and (ii) receptor arrays with heterogeneous sensitivities perform badly.
Our work can thus help to understand how global inhibition shapes normalized odor representations for further processing in the brain.

\end{abstract}

\maketitle

\newlength{\figwidth}
\setlength{\figwidth}{\columnwidth}

\section{Introduction}

Sensory systems encode information efficiently by removing redundancies present in natural stimuli~\citep{Barlow1961, Barlow2001}.
In natural images, for instance, neighboring regions are likely of similar brightness and the  image can thus be characterized by the regions of brightness changes~\cite{Ruderman1994}.
This structure is exploited by ganglion cells in the retina that respond to brightness gradients by receiving excitatory input from photo receptors in one location and inhibitory input from the surrounding~\cite{Demb2015}.
This typical center-surround inhibition results in neural patterns that represent natural images efficiently~\cite{Carandini2012}.
Similarly, such local inhibition helps separating sound frequencies in the ear and locations touched on the skin~\cite{Isaacson2011}.
Vision, hearing, and touch have in common that their stimulus spaces have a metric for which typical correlations in natural stimuli are local.
Consequently, local inhibition can be used to remove these correlations and reduce the high-dimensional input to a lower-dimensional representation.

The olfactory stimulus space is also high-dimensional, since odors are comprised of many  molecules at different concentrations.
Moreover, the concentrations are also often correlated, \eg, because the molecules originate from the same source.
However, these correlations are not represented by neighboring neurons in the olfactory system, since there is no obvious similarity metric for molecules that could be used to achieve such an arrangement~\cite{Nikolova2003}.
Because the olfactory space lacks such a metric, local inhibition cannot be used to remove correlations to form an efficient representation~\cite{Soucy2009, Murthy2011}.
Consequently, the experimentally discovered inhibition in the olfactory system~\cite{Yokoi1995} likely affects neurons irrespective of their location.
Such global inhibition could for instance normalize the activities by their sum, which has been observed experimentally~\cite{Olsen2010, Roland2016}.
This normalization cannot reduce the correlation structure of odors, but it could help separating the odor composition (what is present?) from the  odor intensity (how much is there?)~\cite{Laurent1999, Cleland2010}.
This separation is useful, since the composition identifies an odor source, while the intensity information is necessary for finding or avoiding it.
However, how global inhibition shapes such a bipartite representation of natural odors is little understood.

\begin{figure*}
	\centerline{
		\includegraphics{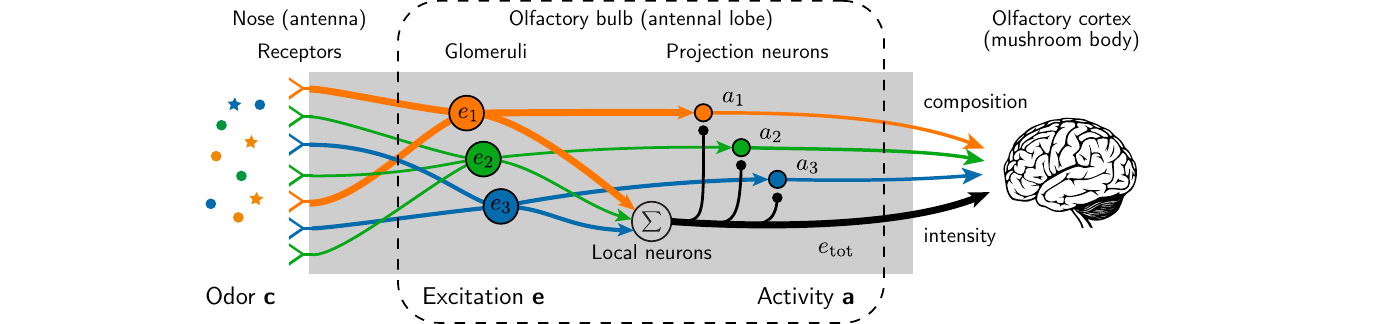}
	}
	\caption{%
	Schematic picture of our model describing the signal processing in the olfactory bulb:
	An odor comprised of many ligands excites the olfactory receptors and the signals from all receptors of the same type are accumulated in respective glomeruli. %
	Associated projections neurons receive excitatory input from a single glomerulus and are subject to global inhibition, mediated by a network of local neurons.
	The activity of the projection neurons form a sparse, concentration-invariant odor representation.
	\label{fig:schematic}
	}
\end{figure*}%

In this paper, we study a simple model of the olfactory system that resembles its first processing layers, which transform the odor representation successively~\cite{Wilson2013, Silva-Teixeira2016}, see \figref{fig:schematic}.
Our model connects previous results from simulations of the neural circuits~\cite{Li1990, Li1994, Linster1997, Getz1999, Cleland2006, Zhang2013} to system-level descriptions of the olfactory system~\cite{Hopfield1999, Koulakov2007, Zwicker2016}.
The main feature of the model is global inhibition, which leads to normalization.
This separates the odor composition from its intensity and encodes it in a sparse representation.
The inhibition strength controls the trade-off between the sparsity and the transmitted information, which influences how well this code can be used to discriminate odors in typical olfactory tasks.
The model reveals two generic consequences of global inhibition:
(i) odors comprised of many different molecules exhibit sparser representations and should thus be more difficult to distinguish and 
(ii) overly sensitive receptors could dominate the sparse responses and arrays with heterogeneous receptors should thus perform poorly.

\section{Simple Model of the Olfactory System}

Odors are blends of odorant molecules that are ligands of the olfactory receptors.
We describe an odor by a vector~$\vect c = (c_1, c_2, \ldots, c_{\Nl})$ that specifies the concentrations~$c_i$ of all $\Nl$ detectable ligands ($c_i \ge 0$).
Generally, only a small subset of the $\Nl \sim 10^5$ ligands are present in natural odors, so most of the $c_i$ will typically be zero.
The ligands in an odor are detected by olfactory receptor neurons, which reside in the nose in mammals and in the antenna in insects~\cite{Kaupp2010}. %
Each of these neurons expresses receptors of one of $\Nr$ genetically defined types, where $\Nr \approx 50$ for flies~\cite{Wilson2013}, $\Nr \approx 300$ for humans~\cite{Verbeurgt2014}, and $\Nr \approx 1000$ for mice~\cite{Niimura2012}.
The excitation of all receptor neurons of the same type is accumulated in associated glomeruli~\cite{Su2009}, whose excitation pattern forms the first odor representation, see \figref{fig:schematic}.
Here, the large number of ligands and their possible mixtures are represented by a combinatorial code, where each ligand typically excites multiple receptor types~\cite{Malnic1999}.
It has been shown experimentally that the excitation~$e_n$ of the glomerulus associated with receptor type~$n$ can be approximated by a linear function of the ligand concentrations~$\vect c$~\cite{Tabor2004, Silbering2007, Gupta2015},
 \begin{align}
	e_n &= \sum_{i=1}^{\Nl} S_{ni} c_i
	\label{eqn:excitation}
	\;,
\end{align}
where $S_{ni}$ denotes the sensitivity of glomerulus~$n$ to ligand~$i$.
We here consider a statistical description of combinatorial coding by studying random sensitivity matrices with entries drawn independently from a log-normal distribution.
This distribution is parameterized by the mean sensitivity~$\bar S$ and the standard deviation~$\lambda$ of the underlying normal distribution.
This choice is motivated by experimental measurements, which also suggest that $\lambda \approx 1$ for flies and humans~\cite{Zwicker2016}.
We showed previously that such random matrices typically decorrelate stimuli and thus lead to near-optimal odor representations on the level of glomeruli~\cite{Zwicker2016}.

In contrast to our previous model, we here consider the odor representation encoded by projection neurons (mitral and tufted cells in mammals), which constitute the next layer after the glomeruli, see \figref{fig:schematic}.
Projection neurons typically receive excitatory input from a single glomerulus~\cite{Jefferis2001} and inhibitory input from many local neurons (granule cells in mammals), which are connected to other projection neurons and glomeruli~\cite{Cleland2010, Su2009}.
The activity~$a_n$ of the projection neurons associated with receptor type~$n$ is a sigmoidal function of ligand concentrations~\cite{Bhandawat2007, Tan2010}.
Additionally, all signals are subject to noise, both from stochastic ligand-receptor interactions and from internal processing~\cite{Lowe1995}, which limits the number of distinguishable output activities.
We capture both effects by considering the simple case where only two activities~$a_n$ can be distinguished.
Here, the projection neurons are active when their excitatory input, the respective excitation~$e_n$, exceeds a threshold~$\gamma$,
\begin{align}
	a_n &= \begin{cases}
		0 & e_n \le \gamma \\
		1 & e_n > \gamma \;.
	\end{cases}
	\label{eqn:activity}
\end{align}
Generally, $\gamma$ could depend on the type~$n$, but we here consider a simple mean-field model, where all types exhibit the same threshold.
Nevertheless, this threshold could still depend on global variables.
Experimental data~\cite{Aungst2003, Silbering2007, Asahina2009, Olsen2010, Hong2015, Banerjee2015, Roland2016, Berck2016} and modeling of the local neurons~\cite{Cleland2006, Cleland2010} suggest that the total excitation of all glomeruli inhibits all projection neurons.
To capture this we postulate that the threshold~$\gamma$ is a function of the total excitation, where we for simplicity consider a linear dependence,
\begin{align}
	\gamma = \frac{\alpha}{\Nr} \sum_{n=1}^{\Nr} e_n
	\label{eqn:threshold}
	\;.
\end{align}
Here, $\alpha$ is a parameter that controls the inhibition strength.

Taken together, our model of the olfactory system comprises~$\Nr$ communication channels, each consisting of receptors, a glomerulus, and projection neurons, which interact via global inhibition, see \figref{fig:schematic}.
The \mbox{\Eqsref{eqn:excitation}--\eqref{eqn:threshold}} describe how this system maps an odor~$\vect c$ to an activity pattern~$\vect a = (a_1, a_2, \ldots, a_{\Nr})$. %
The amount of information that can be learned about~$\vect c$ by observing~$\vect a$ is quantified by the mutual information~$I$, which reads
\begin{align}
	I &= -\sum_{\vect a} P(\vect a) \log_2 P(\vect a)
	\label{eqn:mutual_information}
	\;.
\end{align}
Here, the probability~$P(\vect a)$ of observing output $\vect a$ is given by $P(\vect a) = \int P(\vect a | \vect c) P_{\rm env}(\vect c) \, \diff \vect c$.
The conditional probability $P(\vect a | \vect c)$ of observing~$\vect a$ given~$\vect c$ describes the processing in the olfactory system and follows from the \mbox{\Eqsref{eqn:excitation}--\eqref{eqn:threshold}}.
In contrast, $P_{\rm env}(\vect c)$ denotes the probability of encountering an odor~$\vect c$, which depends on the environment.
Consequently, the information $I$ is not only a function of the sensitivity matrix~$S_{ni}$ and the inhibition strength~$\alpha$, but also of the environment in which the receptors are used~\cite{Zwicker2016}.

Natural odor statistics are hard to measure~\cite{Wright2005} and we thus cannot infer the distribution~$P_{\rm env}(\vect c)$ from experimental data.
Instead, we consider a broad class of distributions parameterized by a few parameters.
For simplicity, we only consider uncorrelated odors, where the concentrations~$c_i$ of ligands are independent.
We denote by $p_i$ the probability that ligand~$i$ is part of an odor.
If this is the case, the associated $c_i$ is drawn from a log-normal distribution with mean~$\mu_i$ and standard deviation~$\sigma_i$.
This choice allows us to independently adjust the mean odor size $s=\sum_i p_i$, the mean of the total concentration~$\ctot=\sum_i c_i$, and the concentration variations~$\frac{\sigma_i}{\mu_i}$.
Averaged over all odors, $c_i$ then has mean $\mean{c_i} = p_i\mu_i$ and variance $\var(c_i) = (p_i - p_i^2) \mu_i^2 + p_i\sigma_i^2$.
Note that typical odors can have hundreds of different ligands~\cite{Wright2005}, but this is still well below~$\Nl \sim 10^5$ and we thus have $1 \ll s \ll \Nl$.

\section{Results}
\subsection{Global inhibition leads to concentration-invariant, sparse representations}

Our model has the interesting property that the odor representation~$\vect a$ does not change when the odor~$\vect c$ or the sensitivities~$S_{ni}$ are scaled by a positive factor.
This is because both the excitations~$e_n$ and the threshold~$\gamma$ are linear in $\vect c$ and $S_{ni}$, see \Eqsref{eqn:excitation} and \eqref{eqn:threshold}, and the activities $a_n$ only depend on the ratio $e_n/\gamma$, see \Eqref{eqn:activity}.
In fact, these equations can be interpreted as normalization of the excitations by the total excitation followed by thresholding with the constant threshold~$\alpha / \Nr$.
Since the representation~$\vect a$ does not depend on $\ctot$, it only encodes relative ligand concentrations, \ie, the odor composition.
This property is called concentration invariance and corresponds to the everyday experiences that odors smell the same over many orders of magnitude in concentration~\cite{Uchida2007, Cleland2011, Zhang2013}.
Indeed, experiments suggest that the activity of projection neurons is concentration-invariant~\cite{Sachse2002,  Sirotin2015, Cleland2007} and exhibits more uniform distances between odors~\cite{Bhandawat2007, Cleland2007}, indicating that they encode the odor composition efficiently.

\begin{figure*}
	\centerline{
		\includegraphics{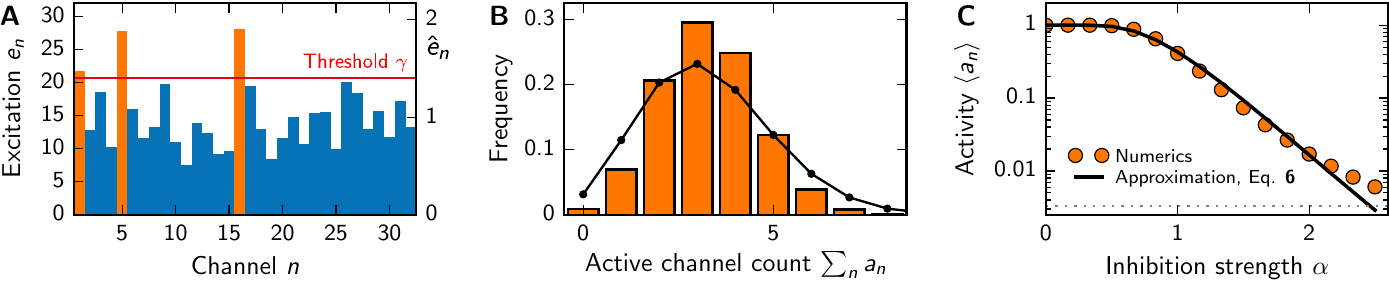}
	}
	\caption{%
	Global inhibition with thresholding leads to sparse odor representations~$\vect a$.
	(A)~Excitations $e_n$ for an arbitrary odor.
	Active channels (orange) have an excitation above the threshold (red line, inhibition strength $\alpha = 1.4$).
	The right axis indicates the normalized excitation $\hat e_n = e_n  \Nr/\sum_m e_m$.
	(B)~Histogram of the number of active channels compared to a binomial distribution (black line) with the same mean for $\alpha = 1.4$.
	(C)~Mean channel activity $\mean{a_n}$ as a function of~$\alpha$.
	The approximation given by \Eqref{eqn:activity_approx} (solid line) is compared to numerical simulations (symbols, standard error of the mean smaller than symbol size).
	The gray dotted line indicates a single expected active channel in humans, $\mean{a_n}=\frac{1}{300}$.
	(A--C) Additional model parameters are $\Nr=32$, $\Nl=256$, $p_i=0.1$, $\mu_i=\sigma_i=1$, and $\lambda=1$. %
	\label{fig:receptor_activity}
	}
\end{figure*}

To understand how odor compositions are encoded in our model, we start with numerical simulations of \Eqsref{eqn:excitation}--\eqref{eqn:threshold} as described in the SI.
\figref{fig:receptor_activity}A shows the excitations $e_n$ corresponding to an arbitrary odor.
Here, the excitation threshold is $1.4$ times the mean excitation, and only three channels are active (orange bars).
The corresponding histogram in \figref{fig:receptor_activity}B shows that the number of active channels is typically small for this inhibition strength when odors are presented with statistics $P_{\rm env}(\vect c)$.
Moreover, the magnitude of the Pearson correlation coefficient between two channels is typically only $\unit[1]{\%}$, see SI.
This weak correlation is expected for the uncorrelated odors and random sensitivity matrices that we consider here and explains why the histogram in \figref{fig:receptor_activity}B is close to a binomial distribution.
The odor representations are thus mainly characterized by the mean channel activity~$\mean{a_n}$.

The mean channel activity~$\mean{a_n}$ depends on the inhibition strength~$\alpha$, the sensitivities~$S_{ni}$, and the odor statistics~$P_{\rm env}(\vect c)$.
To discuss these dependences, we next introduce an approximation based on a statistical description of the associated excitation~$e_n$.
Here, we define the normalized concentrations~$\hat c_i = c_i / \ctot$ and normalized excitations~$\hat e_n = e_n / (\ctot \bar S)$, since $a_n$ is independent of $\ctot$ and $\bar S$.
The statistics of $\hat c_i$ can be estimated in the typical case where odors are comprised of many ligands, see SI.
In the particular case where the ligands are identically distributed the mean is $\mean{\hat c_i} = \Nl^{-1}$
and the variance reads $\var(\hat c_i) \approx (1 - p + \sigma^2 \mu^{-2})/(p\Nl^2)$.
Generally, $\hat c_i$ varies more if the underlying~$c_i$ has higher coefficient of variation~$\sigma_i/\mu_i$ or if the mean odor size~$s=p\Nl$ is smaller.
The normalized excitation~$\hat e_n$ is defined such that its mean is $1$ and the associated variance can be written as a product of the external contribution $V_{\rm ext} = \sum_i \mean{\hat c_i^2}$ due to odors and the internal contribution $V_{\rm int} = \var(S_{ni}) \mean{S_{ni}}^{-2}$ due to sensitivities, see SI.
In the simple case of identically distributed ligands, we have
\begin{align}
	\var(\hat e_n) &= V_{\rm ext} V_{\rm int}
&
	V_{\rm ext} &\approx \frac{1}{s}\left(1+\frac{\sigma^2}{\mu^2}\right)
&
	V_{\rm int} &= e^{\lambda^2} - 1
	\label{eqn:excitations_normalized}
	\;,
\end{align}
for $1 \ll s \ll \Nl$, see SI.
The normalized excitations thus vary more if odors contain fewer ligands, concentrations  fluctuate stronger, or sensitivities are distributed more broadly.
Finally, the mean channel activity~$\mean{a_n}$ is given by the probability that the excitation~$e_n$ exceeds the threshold~$\gamma$, see \Eqref{eqn:activity}.
This is equal to the probability that the normalized excitation~$\hat e_n$ exceeds the normalized threshold~$\hat\gamma = \gamma/(\bar S\ctot)$.
Replacing $\hat\gamma$ by its expectation value $\mean{\hat\gamma}=\alpha$ and using log-normally distributed $e_n$, we obtain
\begin{align}
	\mean{a_n} &\approx \frac{1}{2} \erfc\!\left(
	  \frac{\zeta  + \ln \alpha}
	  {2\zeta^{\frac12}}
	\right)
&\text{with}&&
	\zeta &= \frac12 \ln \bigl(1 + V_{\rm ext}V_{\rm int}\bigr)
	\label{eqn:activity_approx}
\end{align}
for log-normally distributed $\hat e_n$, see SI.
\figref{fig:receptor_activity}C shows that this is a good approximation of the numerical results, which have been obtained from ensemble averages of \Eqref{eqn:activity}. %

The mean activity~$\mean{a_n}$ can also be interpreted as the mean fraction of channels that are activated by an odor, such that small $\mean{a_n}$ corresponds to sparse odor representations.
\figref{fig:receptor_activity}C shows that in our model this is the case for large inhibition strength~$\alpha$, where $\mean{a_n} \sim e^{-\nu}$ with $\nu \approx (\ln\alpha)^2/(4\zeta)$, see SI.
Since sparse representations are thought to be efficient for further processing in the brain~\cite{Laurent1999, Olshausen2004} the inhibition strength~$\alpha$ could be tuned, \eg, on evolutionary time scales, to achieve an activity $\mean{a_n}$ that is optimal for processing the odor representation downstream.
If the optimal value of $\mean{a_n}$ is the same across animals, our theory predicts that inhibition is stronger in systems with more receptor types.
However, this simple argument is not sufficient, since $\mean{a_n}$ also depends on the variations in the natural odor statistics and the receptor sensitivities, which determine $V_{\rm ext}$ and $V_{\rm int}$, respectively.
In particular, the width~$\lambda$ of the sensitivity distribution could also be under evolutionary control.
However, experimental data suggests that both flies and humans exhibit $\lambda\approx 1$~\cite{Zwicker2016}.
Additionally, we show in the SI that much smaller or larger values lead to extremely sparse representations, such that we will only consider $\lambda=1$ in the following.
In this case, the inhibition strength~$\alpha$ controls the sparsity of the odor representation in our simple model of the olfactory system.

\subsection{Sparse coding transmits useful information}

One problem with sparse representations is that they cannot encode as many odors as dense representations.
There is thus a maximal sparsity at which typical olfactory tasks can still be performed.
In general, the performance of the olfactory system can be quantified by the transmitted information~$I$, which is defined in \Eqref{eqn:mutual_information}.
If we for simplicity neglect the small correlations between channels, $I$ can be approximated as~\cite{Zwicker2016}
\begin{align}
	I \approx -\sum_{n=1}^{\Nr} \bigl[
		\mean{a_n} \log_2 \mean{a_n} 
		+ (1 - \mean{a_n})\log_2 (1 - \mean{a_n})
	\bigr]	
	\label{eqn:information_approx}
	\;.
\end{align}
A maximum of $\unit[\Nr]{bits}$ is transmitted when half the channels are active on average, $\mean{a_n}=\frac12$.
In our model, this is the case for weak inhibition, $\alpha < 1$, see \figref{fig:receptor_activity}C.
In the opposite case of significant inhibition, $\alpha > 1$,  few channels are typically active and the transmitted information is smaller.
In the limit $\mean{a_n} \ll 1$, the information is approximately given by $I \sim \frac{1}{\ln 2} \Nr\mean{a_n} \cdot (1 - \ln\mean{a_n})$, which implies that even if only $\unit[10]{\%}$ of the channels are active on average, the information~$I$ is still almost half of the maximal value of $\unit[\Nr]{bits}$.
However, large information~$I$ does not automatically indicate a good receptor array, since only accessible information that can be used to solve a given task matters~\cite{Tikhonov2015, Tkacik2016}.

To test whether sparse representations are sufficient to solve typical olfactory tasks, we next study how well odors can be discriminated in our model.
As a proxy for the discriminability, we calculate the Hamming distance~$d$ between the odor representations, which is given by the number of channels with different activity.
In the simple case of uncorrelated odors, which do not share any ligands, the expected distance~$\mean d$ is approximately given by total number of active channels in both representations.
Consequently, uncorrelated odors can be distinguished even if their representations are very sparse.
However, realistic tasks typically require distinguishing similar odors.
We thus next study the discriminability of odors that vary in the relative concentrations of their ligands, their size, and their composition.

\begin{figure*}
	\centerline{
		\includegraphics{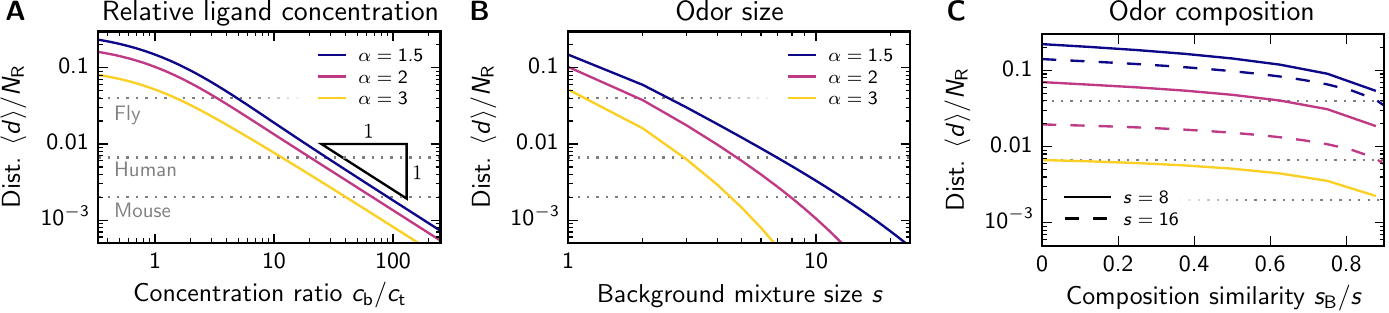}
	}
	\caption{%
	Sparse coding is sufficient to distinguish odors with different relative ligand concentrations, size, and composition.
	(A)~Mean distance $\mean{d}$ between the representations of a background ligand at concentration $\cb$ and an odor with an additional target ligand at concentration $\ct$ as a function of the dilution~$\cb/\ct$ for various inhibition strengths~$\alpha$.	
	(B)~Distance $\mean{d}$ resulting from adding a ligand to an odor comprised of $s$ ligands as a function of~$s$ for various~$\alpha$. %
	(C) Distance $\mean{d}$ between the representations of two odors with $s$ ligands, sharing $\sB$ of them, as a function of the similarity $\sB/s$ for small ($s=8$, solid lines) and large odors ($s=16$, dashed lines).
	The colors indicate the same~$\alpha$ as in the other panels.
	(A--C)~The gray dotted lines indicate the threshold $\mean{d}=2$ for $\Nr=50, 300, 1000$ (corresponding to flies, humans, and mice; top to bottom). 
	The width of the sensitivity distribution is $\lambda=1$.
	\label{fig:odor_discriminability}
	}
\end{figure*}%

We start by determining the maximal dilution~$\frac\cb\ct$ at which a target odor at concentration $\ct$ can still be detected in a background of concentration~$\cb$.
We calculate the expected difference~$\mean{d}$ between the associated representations from the probability that a given channel changes its activity when the target is added, see SI.
Since this probability is the same for all channels, $\mean{d}$ is proportional to the number~$\Nr$ of channels.
For the simple case where both the target and the background are a single ligand, \figref{fig:odor_discriminability}A shows that $\mean{d}$ decreases for smaller target concentrations and is qualitatively the same for all inhibition strengths~$\alpha$.
For large dilutions~$\frac\cb\ct$, $\mean{d}$ is inversely proportional to the dilution, $\mean{d} \propto \Nr \frac\ct\cb$.
Since the addition of the target can only be detected reliably if $\mean d > 2$, which corresponds to a situation where one channel becomes inactive and another one active, our model predicts that doubling the number~$\Nr$ of channels also doubles the concentration sensitivity.
\figref{fig:odor_discriminability}A thus implies that mice ($\Nr\approx 1000$) should be able to detect the addition of a target even if it is almost a hundred times more dilute than the background, which is close to the threshold that has been found experimentally~\cite{Mouret2009}.
Conversely, flies ($\Nr \approx 50$) should fail for very small dilution factors.

We next study odors comprised of many ligands, since typical odors are blends~\cite{Wright2005}.
For simplicity, we consider the detection of a single target ligand in a background mixture of varying size~$s$ when the target ligand and the ligands in the background have equal concentration, such that the target dilution is $s$.
\figref{fig:odor_discriminability}B shows that the qualitative dependence of $\mean d$ on the dilution is similar to the single ligand case in panel A, but the maximal dilution for detecting the target is different.
For instance, the model predicts that mice cannot identify the addition of the target ligand to a background consisting of more than ten ligands, while the maximal dilution was almost one hundred in the case of single background ligands.
Consequently, the discrimination performance seems to drop significantly when larger odors are considered.
This qualitatively agrees with experiments where humans are not able to identify all ligands in mixtures of more than three ligands \cite{Jinks2001, Goyert2007} and they fail to detect the presence or absence of ligands in mixtures of more then $15$ ligands~\cite{Jinks1999}.
Even if humans cannot identify ligands in large odors, they might still be able to distinguish two such odors.
To study this, we next compare the representations of two odors that each contain $s$ ligands, sharing $\sB$ of them, for the simple case where all ligands have the same concentration.
\figref{fig:odor_discriminability}C shows that the distance~$\mean d$ between the two odors decreases with larger $\sB$, \ie, more similar odors are more difficult to discriminate.
However, $\sB$ only has a strong effect if more than about $\unit[80]{\%}$ of the ligands are shared between odors.
Conversely, the inhibition strength~$\alpha$ and the odor size~$s$ significantly influence~$\mean d$ for all values of $\sB$.
This agrees with the results shown in \figref{fig:odor_discriminability}B, where $\mean d$ exhibits a similar dependence on $\alpha$ and $s$.
While it is expected that the performance decreases with large inhibition strength~$\alpha$ since fewer channels are active, the strong dependence on the size~$s$ is surprising.

\subsection{Larger odors have sparser representations}
Why are odors with many ligands more difficult to discriminate in our model?
Since correlations between channels seem to be negligible, the most likely explanation is that larger odors activate fewer channels.
To test this hypothesis, we determine the activity~$\mean{a_n}$ in the simple case where all ligands in an odor have the same concentration.
Because of the normalization, the value of this concentration does not matter and $\mean{a_n}$ only depends on the inhibition strength $\alpha$ and the odor size $s$.
In the limit of large odors ($s \gg 1$), the approximation given in \Eqref{eqn:activity_approx} yields $\mean{a_n} \sim e^{-\beta s}$ with $\beta \sim (\ln\alpha)^2$, see SI.
In this case, the activity $\mean{a_n}$ thus decreases exponentially with $s$ and this decrease is stronger for larger $\alpha$.
Consequently, larger odors activate fewer channels and it is thus less likely that a small change in such odors alters the activation pattern~$\vect a$.

\begin{figure*}
	\centerline{
		\includegraphics{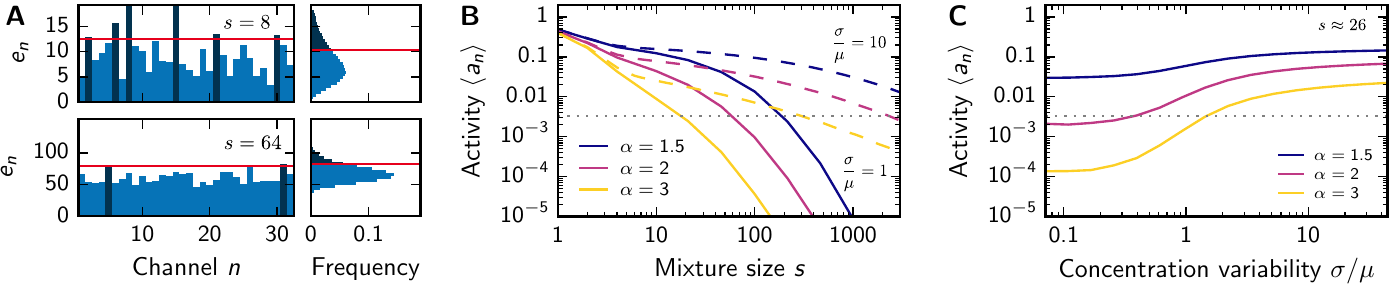}
	}
	\caption{%
	Larger odors activate fewer channels.
	(A) Comparison of the excitations~$e_n$ of small (odor size $s=8$, upper panels) and large odors ($s=64$, lower panels) at $\alpha=1.3$.
		$e_n$ for a single odor (left panels) and histograms for all odors (right panels) are shown.
		Larger odors exhibit fewer active channels (dark blue), for which the excitations are above threshold (red line).
	(B) Numerically determined~$\mean{a_n}$ as a function of $s$ for various inhibition strengths~$\alpha$ at small ($\sigma/\mu=1$, solid lines) and large concentration variability ($\sigma/\mu=10$, dashed lines) at $\Nl=10^4$.
	(C) Numerically determined $\mean{a_n}$ as a function of $\sigma/\mu$ for various~$\alpha$.
	(A--C) Additional model parameters are $\Nr=32$, $\Nl = 256$, $p_i = 0.1$, $\mu_i=\sigma_i=1$, and $\lambda=1$.
	The gray dotted line in B and C indicates a single expected active channel in humans, $\mean{a_n}=\frac{1}{300}$.
	\label{fig:receptor_activity_measurement}
	}
\end{figure*}%

Larger odors activate fewer channels because the respective excitations~$e_n$ have a smaller variability.
For an odor with $s$ ligands of equal concentration, $e_n$ is proportional to the sum of $s$ sensitivities~$S_{ni}$, see \Eqref{eqn:excitation}.
Consequently, $e_n$ can be considered as a random variable whose mean $\mean{e_n}$ and variance $\var(e_n)$ scale with $s$.
The activity~$\mean{a_n}$ is  given by the fraction of excitations that exceed the threshold~$\gamma$, which also scales with~$s$.
This fraction typically scales with the coefficient of variation $\var(e_n)^{\frac12}\mean{e_n}^{-1}$, which is proportional to $s^{-\frac12}$ and is thus smaller for larger odors.
Larger odors thus activate fewer channels because there are fewer excitations that are much larger than the mean, see \figref{fig:receptor_activity_measurement}A.
This is a direct consequence of the assumption that the excitation threshold~$\gamma$ scales with the mean excitation and this result does not depend on other details of the model.
Conversely, the dependence of $\mean{a_n}$ on the inhibition strength~$\alpha$ is model specific, since it follows from the shape of the tail of the excitation distribution.
In particular, the influence of the odor size on $\mean{a_n}$ is insignificant for weak inhibition, $\alpha \approx 1$, because approximately half the channels are activated irrespective of the variance~$\var(e_n)$.

This qualitative explanation illustrates that depending on the variability of the excitations different odors can have representations with very different sparsities.
Indeed, we find that the sparsity changes over several orders of magnitude as a function of the odor size~$s$ in our model, see \figref{fig:receptor_activity_measurement}B.
Moreover, the concentration variability~$\frac\sigma\mu$ of the individual ligands also has a strong effect on the sparsity, see \figref{fig:receptor_activity_measurement}C.
This is because larger $\frac\sigma\mu$ implies larger variations in the excitations, such that more channels exceed the threshold and become active.
In fact, this dependence of $\mean{a_n}$ on $s$ and $\frac\sigma\mu$ is also qualitatively captured by the analytical approximation given in \Eqref{eqn:activity_approx}, which explicitly depends on the odor variability~$V_{\rm ext}$ defined in \Eqref{eqn:excitations_normalized}.
Taken together, our model shows that the sparsity of the odor representations strongly depend  on the odor statistics~$P_{\rm env}(\vect c)$.

\subsection{Effective arrays have similar receptor sensitivities}
\begin{figure*}
	\centerline{
		\includegraphics{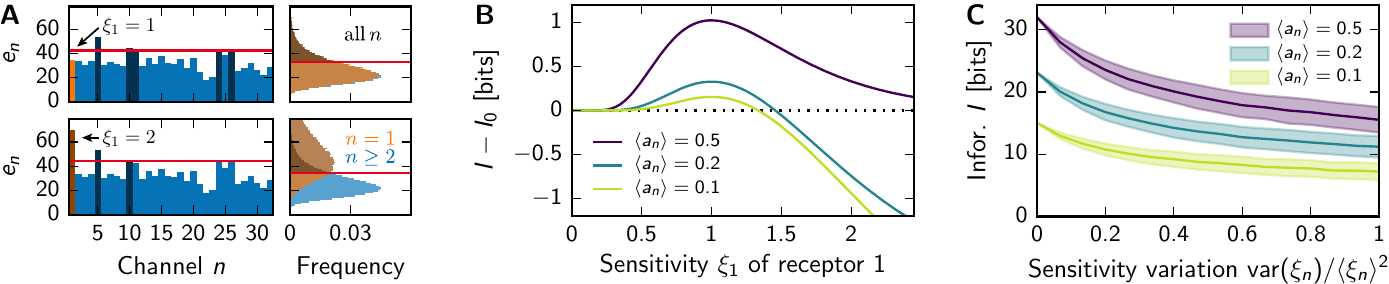}
	}
	\caption{%
	Receptors with heterogeneous sensitivities make poor arrays.
	(A)~Comparison of the excitations~$e_n$ for homogeneous ($\xi_1=1$, upper panels) and heterogeneous receptors ($\xi_1=2$, lower panels).
		$e_n$ for the same arbitrary odor (left panels) and histograms for all odors (right panels) are shown for the first receptor ($n=1$, orange) and all other receptors ($n\ge 2$, blue).
		Dark bars indicate excitations that are above the threshold (red line, inhibition strength~$\alpha=1.3$).
	(B)~Information~$I$ given by \Eqref{eqn:information_approx} as a function of the sensitivity $\xi_1$ of the first receptor. %
	The channel activity $\mean{a_n}$ calculated from \Eqref{eqn:activity_receptor_factors} is set to the given value by adjusting~$\alpha$.
	$I$ is shown relative to the information~$I_0$ of a system without the first receptor (dotted line).
	(C)~Information~$I$ (line, mean; shaded area indicates standard deviation) of log-normally distributed~$\xi_n$ as a function of the variation~$\var(\xi_n)\mean{\xi_n}^{-2}$ for various~$\mean{a_n}$.
	(A--C) 
	Remaining parameters are $\Nr=32$, $\Nl=256$, $p_i=0.1$, $\mu_i=\sigma_i=1$, and $\lambda=1$.
	\label{fig:receptor_factors}
	}
\end{figure*}

So far, we considered homogeneous receptor arrays, where all receptor types have the same average sensitivity.
However, realistic receptors vary in their biochemical details and it might thus be difficult to have such homogeneous arrays.
We thus next consider the effect of sensitivity variations between different receptors.
This is important, since a channel with overly sensitive receptors will contribute significantly to the common threshold~$\gamma$, suppress the activity of other channels, and could thus limit the coding capacity of the system, see \figref{fig:receptor_factors}A.
To study this, we consider sensitivity matrices~$S_{ni} = \xi_n S^{\rm iid}_{ni}$, where $\xi_n$ denotes the mean sensitivity of receptor type~$n$ and $S^{\rm iid}_{ni}$ is the sensitivity matrix that we discussed so far, \ie, it is a random matrix where all entries are independently drawn from a log-normal distribution described by the mean~$\bar S$ and width~$\lambda$.
Here, $\xi_n$ captures differences between receptor types, \eg, because of biochemical differences or due to variations in copy number, see SI.
For this model, the mean excitation threshold is $\mean{\gamma} = \alpha\bar S\mean{\ctot} \xitot\Nr^{-1}$ where $\xitot = \sum_n \xi_n$.
The expected channel activity is approximately given by
\begin{align}
	\mean{a_n} &\approx 1 - F\biggl( 
	    \frac{\alpha \xitot}{N_{\rm r} \xi_n}
    \biggr)
	\;,
	\label{eqn:activity_receptor_factors}
\end{align}
where $F(\hat e_n)$ is the cumulative distribution function of the normalized excitations~$\hat e_n$ for $\xi_n=1$, whose mean is $\mean{\hat e_n}=1$ and whose variance is given by~\Eqref{eqn:excitations_normalized}.
Note that~$\mean{a_n}$ does not change if all $\xi_n$ are multiplied by the same factor.
In particular, the expression above reduces to $\mean{a_n} \approx 1 - F(\alpha)$ and thus \Eqref{eqn:activity_approx} if all $\xi_n$ are equal.

We first discuss the influence of the receptor sensitivities~$\xi_n$ by only varying one type, \ie, we change $\xi_1$ while setting $\xi_n=1$ for $n\ge 2$.
\figref{fig:receptor_factors}B shows that for fixed channel activity~$\mean{a_n}$ the transmitted information~$I$ is maximal for a homogeneous receptor array (\mbox{$\xi_1=1$}).
$I$~is reduced for smaller~$\xi_1$ and for $\xi_1=0$ it reaches the value~$I_0$ of an array where the first receptor was removed.
Conversely, $I$ can drop well below $I_0$ when $\xi_1$ is increased above $1$.
In this case, the large excitation of the affected channel not only leads to its likely activation, but it also raises the threshold~$\gamma$ and thereby inhibits other channels, see \figref{fig:receptor_factors}A.
In the extreme case of very large $\xi_1$, this channel will always be active while all other channels are silenced, which implies $I=0$.
There is thus a critical value of~$\xi_1$ beyond which removing the receptor from the array is advantageous for the overall performance.
\figref{fig:receptor_factors}B shows that increasing the sensitivity of a receptor by only $\unit[40]{\%}$ can make it useless in the context of the whole array if representations are sparse.

So far, we only varied the sensitivity of a single receptor. 
To test how variations in the sensitivities of all receptors affect the information~$I$, we next consider log-normally distributed~$\xi_n$.
Here, vanishing variance of $\xi_n$ corresponds to a homogeneous receptor array.
\figref{fig:receptor_factors}C shows that small variations in $\xi_n$ can strongly reduce the transmitted information~$I$.
Since~$I$ limits the discriminative capability of the receptor array, this suggests that receptor arrays with heterogeneous sensitivities perform worse.

The simple model that we discuss here shows that the excitation statistics of the different channels determine the properties of the resulting odor representation.
In particular, receptors that have lower excitations on average might be suppressed often and thus contribute less to the odor information.
Since the excitation statistics are influenced both by the sensitivities~$S_{ni}$ and the odor statistics~$P_{\rm env}(\vect c)$, this suggests that the sensitivities should be adjusted to the odor statistics.
In an optimal receptor array, the sensitivities are chosen such that all channels have the same probability to become active.

\section{Discussion}

We studied a simple model of odor representations, which is based on normalization and a non-linear gain function.
This model separates the odor composition, encoded in the activity~$\vect a$ of the projection neurons, from the odor intensity, which could be encoded by the total excitation~$e_{\rm tot}$ or the threshold level~$\gamma$~\cite{Mainland2014}.
For significant inhibition the representation~$\vect a$ is sparse and the set of active projection neurons provides a natural odor 'tag' that could be used for identification and memorization in the downstream processing~\cite{Stevens2015}.

Sparse representations reduce the coding capacity and transmit less information than dense ones.
However, even if the mean activity is $\mean{a_n} = 0.01$ and thus $50$ times smaller than in maximally informative arrays with $\mean{a_n} = 0.5$, the transmitted information~$I$ is only reduced by a factor of $12$, see \Eqref{eqn:information_approx}.
For humans with $\Nr=300$, this yields $I \approx \unit[25]{bits}$, allowing to encode $2^I \approx 10^7$ different odor compositions.
Note that the total information~$I_{\rm tot}$ also includes information~$I_{\rm int}$ about the odor intensity, $I_{\rm tot} = I + I_{\rm int}$.
Here, $I_{\rm int} \approx \unit[10]{bits}$ would be sufficient to encode the total concentration over a range of $10$ orders of magnitude with a resolution of $\unit[5]{\%}$, typical for humans~\cite{Cain1977}.
In this case, our model compresses the $\unit[300]{bits}$ of a maximally informative representation on the level of glomeruli~\cite{Zwicker2016} to only $I_{\rm tot} \approx \unit[35]{bits}$ on the level of projection neurons.

The model discussed here is similar to our previous model, where we discussed representations on the level of the glomeruli~\cite{Zwicker2016}.
Both models use a maximum entropy principle to determine properties of optimal receptor arrays.
To achieve this, the receptor sensitivities must be tailored to the odor statistics in both models.
The main difference of the models is the global inhibition discussed here, which separates the odor composition from its intensity and thus removes the correlation between the glomeruli excitation and the odor intensity~\cite{Haddad2010}.
Consequently, odors can then be discriminated at all concentrations, while this was only possible in a narrow concentration range in the glomeruli model~\cite{Zwicker2016}.
The additional normalization is thus useful to separate odors, even if the projection neurons encode less information than the respective glomeruli.
To estimate this information, we consider binary outputs in both models, which corresponds to very noisy channels.
However, the glomeruli model discusses arrays of noisy receptor, while we here consider perfect receptors whose signal is first normalized and then subjected to noise.
This additional processing reduces correlations and leads to sparse representations, which might simplify downstream computations.
Consequently, this model is suitable for describing natural olfaction, where the capacity for the downstream computations is limited, while the glomeruli model is relevant for artificial olfaction~\cite{Stitzel2011}, since computers have enough power to handle high-dimensional signals.

Sparse responses of projection neurons have been observed in experiments~\cite{Davison2007, Rinberg2006}.
For instance, in mice $\unit[15]{\%}$ of the projection neurons respond to a given single ligand~\cite{Roland2016}, suggesting significant inhibition.
However, in locust about two third of the projection neurons respond to any given odor~\cite{Perez-Orive2002}, which implies weak inhibition.
It is thus conceivable that some animals exhibit sparse representations while others have maximally informative ones, although additional experiments are needed to characterize the representations better.
A direct experiment could test whether the odor percept changes when the weakly responding glomeruli are disabled artificially.
Additionally, it will be important to study the representations of mono-molecular odors and mixtures at various concentration to better resemble the natural odor statistics.
For instance, our simple theory predicts that fewer than $\unit[15]{\%}$ of the projection neurons in mice respond when complex mixtures are presented.
Indeed, experiments find that only $3$ to $\unit[10]{\%}$ of the projection neurons in mice fire for complex urine odors~\cite{Lin2005}.
Conversely, the statistics of the activity of projection neurons in flies seem to be independent of the stimulus~\cite{Stevens2016}.
Our theory can also be tested by measuring how well odors can be discriminated.
For instance, odors are much more difficult to distinguish if they contain more ligands in our model, which has also been observed experimentally~\cite{Weiss2012}.
Conversely, other experiments indicate that the odor size only weakly influences the odor discriminability~\cite{Bushdid2014}.
Taken together, there is some experimental evidence that the odor representations and thus the discriminability change with odor size, although there is also evidence to the contrary, which could hint at mechanisms beyond global inhibition that influence the odor representations.

The coding sparsity given by the mean channel activity~$\mean{a_n}$ can be adjusted by changing the inhibition strength~$\alpha$ or the width~$\lambda$ of the receptor sensitivity distribution in our model.
Additionally, $\mean{a_n}$ is a function of the natural odor statistics, \ie, the typical number of ligands in odors and their concentration distribution.
Consequently, $\alpha$ or $\lambda$ must be adjusted to keep $\mean{a_n}$ constant if the odor statistics change, \eg, because of seasonal changes or migration to a different environment.
This adjustment could happen on multiple timescales, reaching from evolutionary adaptations of the receptors to near-instantaneous adjustments of the involved neurons, and it is likely that the global inhibition is regulated on all levels~\cite{Wilson2013}.
In this paper, we investigated the simple case of constant~$\alpha$ and $\lambda$, which corresponds to slow regulation, but it is conceivable that $\alpha$ could be regulated on short time scales.
For instance, the threshold could be lowered for larger odors to improve their discriminability.
Our model suggests that such additional mechanisms are necessary to efficiently discriminate odors of all sizes.

Our model also reveals that it is important to control the properties of the individual communication channels to have useful receptor arrays.
For instance, increasing the sensitivity of a given receptor by $\unit[40]{\%}$ can be worse then removing it completely, see \figref{fig:receptor_factors}A.
Generally, a receptor array is only effective if the different channels have similar excitations on average.
This suggests that the sensitivities are tightly controlled and maybe even adjusted to the odor statistics of the environment.
On evolutionary time scales, the sensitivities could be regulated by point mutations of the receptors that change how ligands bind~\cite{Yu2015}.
On shorter time scales, the sensitivities could be regulated by changing the receptor copy numbers, see SI.
Since this is observed experimentally~\cite{Yu2016}, we predict that the receptor copy numbers are adjusted such that the excitations of all glomeruli are similar when averaged over natural odors.
Alternatively, variations in the receptor sensitivities could be balanced by more complex inhibition mechanism.
For instance, experiments show that different projection neurons have different susceptibilities to inhibition~\cite{Hong2015}.
Here, the experimentally observed turnover of mitral cells and interneurons~\cite{Lazarini2011} could adjust the inhibition mechanism locally, which could optimize the olfactory system for a given environment~\cite{Mouret2009}.
Such adaptation of the inhibition mechanism to the current stimulus statistics and more complex models where the behavioral state of an animal could influence the olfactory bulb by top-down modulation~\cite{Wilson2013} will be interesting to explorer in the future.

Our simplified model neglects many details of the olfactory system~\cite{Silva-Teixeira2016}.
For instance, we do not consider the dynamics of inhalation and the odor absorption in the mucus~\cite{Pelosi2001, Schoenfeld2005}.
Instead, we here directly parameterize the ligand distribution at the olfactory receptors, where we for simplicity neglect correlations between ligands.
It would be interesting to extend the model for more complex stimuli and study how the system decorrelates the input, identifies a target odor in a background, and separates multiple odors from each other.
This likely involves many steps~\cite{Cleland2011} and cannot be done perfectly with a single  normalization step and non-linear gain function.
For instance, it might be important to apply gain functions at the level of receptors and the glomeruli to model finite sensitivity and saturation effects.
Additionally, it has been shown that there is additional cross-talk on the level of receptors~\cite{Ukhanov2010} and glomeruli~\cite{Aungst2003, Silbering2007}, which could support decorrelation.
Generally, such cross-talk and the inhibition that we discussed here will be non-linear~\cite{Wilson2011a}.
This could for instance be modeled by a divisive normalization model that has been proposed for olfaction~\cite{Olsen2010}.
It is also likely that the inhibition of the projection neurons is not driven by a single global variable.
If glomeruli positioning carried some meaning~\cite{Murthy2011}, local inhibition could  help separating similar odors by enhancing the contrast~\cite{Leon2003}.
The discrimination of similar odors could also be improved if projection neurons had a larger output range, increasing the information capacity per channel.
Finally, we completely neglected the temporal dynamics of the olfactory system, which play an important role for the adaptation between sniffs~\cite{Zufall2000} and might also influence odor perception within a single sniff~\cite{Blauvelt2013, Sirotin2015, Uchida2014}.

\begin{acknowledgments}
I thank Michael P. Brenner, Venkatesh N. Murthy, Mikhail Tikhonov and Christoph A. Weber for helpful discussions and a critical reading of the manuscript.
This research was funded by the Simons Foundation and the German Science Foundation through ZW \mbox{222/1-1}.
\end{acknowledgments}

\bibliographystyle{aipauth4-1}
\bibliography{bibdesk}

\end{document}


\title{Supporting Information: Olfactory coding with global inhibition}
\title{Supporting Information: Normalized neural representations of natural odors}

\date{\today}

\author{David Zwicker}
\email{dzwicker@seas.harvard.edu}
\homepage{http://www.david-zwicker.de}
\affiliation{School of Engineering and Applied Sciences, Harvard University, Cambridge, MA 02138, USA}
\affiliation{Kavli Institute for Bionano Science and Technology, Harvard University, Cambridge, MA 02138, USA}

\maketitle
\tableofcontents

\section{Statistics of normalized concentrations and excitations}
Let $p_i$ be the probability that ligand~$i$ is present in an odor.
If it is present, its concentration~$c_i$ is drawn from a log-normal distribution with mean~$\mu_i$ and standard deviation~$\sigma_i$, while $c_i=0$ if the ligand is not present.
Hence,
\begin{subequations}
\begin{align}
	\mean{c_i} &= p_i\mu_i
\\
	\var(c_i) &= (p_i - p_i^2) \mu_i^2 + p_i\sigma_i^2
	\;,
\end{align}
\end{subequations}
while the covariances~$\cov(c_i, c_j) = \mean{c_ic_j} - \mean{c_i}\mean{c_j}$ vanish for $i\neq j$ since the ligands are independent.
The statistics of the total concentration~$\ctot = \sum_i c_i$ read
\begin{align}
	\mean{\ctot} &= \sumi \mean{c_i}
& \text{and} && 
	\var(\ctot) &= \sumi  \var(c_i)
	\;.
	\label{eqn:ctot_stats}
\end{align}
The excitations~$e_n$ are given by $e_n = \sum_i S_{ni} c_i$, where the sensitivities~$S_{ni}$ are log-normally distributed with mean $\mean{S_{ni}} = \bar S$ and variance $\var(S_{ni}) = \bar S^2(e^{\lambda^2} - 1)$.
Hence,
\begin{subequations}
\begin{align}
	\mean{e_n} &= \bar S \mean\ctot
\\
	\var(e_n) &= \bar S^2 \var(\ctot)
		+ \var(S_{ni}) \sumi \mean{c_i^2}
	\;,
\end{align}
\end{subequations}
where $\mean{c_i^2} = p_i(\mu_i^2 + \sigma_i^2)$ and $\cov(e_n, e_m)=0$ for $n\neq m$.

We next determine the statistics of the normalized concentrations~$\hat c_i = c_i/\ctot$.
For simplicity, we consider large odors, $\sum_i p_i \gg 1$, where $\ctot$ can be considered as an independent random variable.
Since $\ctot$ is the sum of (a variable) number of log-normally distributed random variables, its distribution can be approximated by another log-normal distribution~\cite{Wu2005}, which we parameterize by its mean~$\mu_{\rm tot}$ and variance~$\sigma_{\rm tot}^2$.
We consider the simple approximation where these parameters are directly given by \Eqref{eqn:ctot_stats}~\cite{Fenton1960}.
This choice approximates the tail of the distribution well, but leads to errors in the vicinity of the mean~\cite{Wu2005}.

Since both $\ctot$ and $c_i$ are log-normally distributed when ligand~$i$ is present in an odor ($c_i>0$), $\hat c_i$ is also log-normally distributed in this case and 
\begin{subequations}
\begin{align}
	\mean{\hat c_i}_{c_i > 0} &= \frac{\mu_i}{\mu_{\rm tot}} \chi
\\
	\var(\hat c_i)_{c_i > 0} &= 
		\frac{\mu_i^2\chi^2}{\mu_{\rm tot}^2}
  		\left(\frac{\sigma_i^2}{\mu_i^2} \chi
		+ \chi - 1
	    \right)
	 \;,
\end{align}
\end{subequations}
where $\chi = 1 + \sigma_{\rm tot}^2 \mu_{\rm tot}^{-2}$.
Since $\hat c_i=0$ with probability $1-p_i$, the statistics of $\hat c_i$ read
\begin{subequations}
\label{eqn:norm_conc_stats}
\begin{align}
	\mean{\hat c_i} &= \frac{p_i\mu_i}{\mu_{\rm tot}} \chi
	\label{eqn:norm_conc_mean}
\\
	\var(\hat c_i) &= 
		\frac{p_i\mu_i^2\chi^2}{\mu_{\rm tot}^2}\left(
			\frac{\sigma_i^2}{\mu_i^2}\chi
			+ \chi  - p_i
  	  \right)
	\;.
\end{align}
\end{subequations}
Note that the covariance~$\cov(\hat c_i, \hat c_j)$ does not vanish since the $\hat c_i$ are not independent.
In particular, $\var(\sum_i \hat c_i)=0$, since $\sum_i \hat c_i = 1$ by definition.
This condition is only consistent with $\Eqref{eqn:norm_conc_mean}$ if $\chi \approx 1$, which implies that $\ctot$ must not vary much, $\frac{\sigma_{\rm tot}}{\mu_{\rm tot}} \ll 1$.
Using $\chi = 1$, the statistics of the normalized excitations~$\hat e_n = \bar S^{-1} \sum_i S_{ni} \hat c_i$ read
\begin{subequations}
\begin{align}
	\mean{\hat e_n} &= 1 %
\\
	\var(\hat e_n) &= 
		\frac{\var(S_{ni})}{\bar S^2} \Bigl\langle\sum_i \hat c_i^2 \Bigr\rangle
	  \;,
\end{align}
\end{subequations}
where $\mean{\sum_i \hat c_i^2} \approx \sum_i\mean{\hat c_i^2}$ with $\mean{\hat c_i^2} = \mean{\hat c_i}^2 + \var(\hat c_i)$ and the statistics given in \Eqref{eqn:norm_conc_stats}.

In the simple case where all ligands are drawn from the same distribution ($p_i = p$, $\mu_i = \mu$, $\sigma_i=\sigma$), we obtain 
\begin{align}
	\mean{\hat c_i} &\approx \frac1\Nl \;,
	&
	\var(\hat c_i) &\approx  
		 \frac{1-p+\frac{\sigma^2}{\mu^2}}{s\Nl}
		\;,
\end{align}
and $\mean{\hat c_i^2}  \approx \frac{1}{s\Nl}(\frac{\sigma^2}{\mu^2} + 1)$,
such that
\begin{align}
	\var(\hat e_n) & \approx 
		\frac{1}{s}
		\left( 1 + \frac{\sigma^2}{\mu^2} \right)
		\frac{\var(S_{ni})}{\bar S^2}
	\;,
\end{align}
which is equivalent to Eq.~{\bf 5} in the main text.

\section{Numerical simulations}

\begin{figure}
	\centerline{
		\includegraphics[width=\columnwidth]{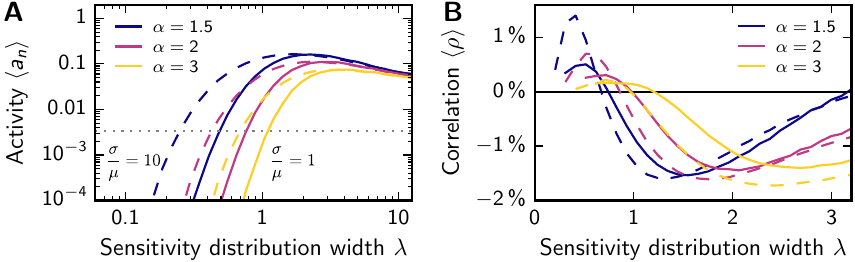}
	}
	\caption{%
	Influence of the width~$\lambda$ of the sensitivity distribution on the statistics of the odor representations.
	(A)~Expected channel activity~$\mean{a_n}$ as a function of~$\lambda$ for several inhibition strengths~$\alpha$.
	Intermediated values, $\lambda \approx 1$, lead to larger activities.
	(B)~Mean Pearson correlation coefficient~$\mean\rho$ calculated from an ensemble average of \Eqref{eqn:pearson} as a function of~$\lambda$ for several~$\alpha$.
	For small~$\lambda$, $\mean{a_n}$ was too small to estimate~$\rho$ reliably.
	(A--B)~Results are shown for small ($\frac\sigma\mu=1$, solid lines) and large ($\frac\sigma\mu=10$, dashed lines) concentration variability.
	Remaining parameters are $\Nr=32$, $\Nl=256$, and $p_i=0.1$.
	\label{fig:activity_width}
	}
\end{figure}

We numerically calculated ensemble averages over odors~$\vect c$ and sensitivity matrices~$S_{ni}$.
Here, we first choose $S_{ni}$ by drawing all entries independently from a log-normal distribution with mean~$\bar S=1$ and variance~$\var(S_{ni}) = e^{\lambda^2} - 1$.
We then draw an odor~$\vect c$ using the following procedure:
First, we determine which of the~$\Nl$ ligands are present according to their probabilities~$p_i$.
Second, we draw the concentrations~$c_i$ for each ligand~$i$ that is present from a log-normal distribution with mean~$\mu_i$ and standard deviation~$\sigma_i$.
We then use Eqs.~{\bf 1}--{\bf 3} given in the main text to map the odor~$\vect c$ to a binary activity vector~$\vect a$, from which we can for instance calculate the number of active channels.
We obtain ensemble averages of such quantities by repeating these steps $10^5$ times.
This allows us to calculate the mean activities~$\mean{a_n}$, the covariances~$\cov(a_n, a_m)$, and the  Pearson correlation coefficient~$\rho$, which is defined as
\begin{align}
	\rho = \frac{1}{\Nr^2 - \Nr} \sum_{n \neq m} 
		\frac{\cov(a_n, a_m)}{\bigl[\var(a_n)\var(a_m)\bigr]^{\frac12}}
	\;.
	\label{eqn:pearson}
\end{align}
\figref{fig:activity_width} shows these quantities as a function of the width~$\lambda$ of the sensitivity distribution.
We also estimate~$P(\vect a)$ from an ensemble average to calculate the information~$I$ from its definition given in Eq.~{\bf 4} in the main text.

\section{Approximate channel activity}
We estimate the expected activity~$\mean{a_n}$ by the probability that the normalized excitations~$\hat e_n$ exceed the expected normalized threshold~$\alpha$.
Since both the sensitivities~$S_{ni}$ and the normalized concentrations~$\hat c_i$ are approximately log-normally distributed, $\hat e_n$ can also be approximated by a log-normal distribution~\cite{Fenton1960}.
The associated probability distribution function reads
\begin{align}
	f(\hat e_n) &= 
		\frac{1}{\sqrt{2 \pi } S_n \hat e_n}
			\exp\left[-\frac{\bigl(M_n - \ln (\hat e_n)\bigr)^2}{2 S_n^2}\right]
	\label{eqn:lognorm_pdf}
\end{align}
and the cumulative distribution function is
\begin{align}
	F(\hat e_n) &=\frac12 \erfc\left[
		\frac{M_n - \ln(\hat e_n)}{\sqrt 2  S_n}
	\right]
	\label{eqn:lognorm_cdf}
	\;.
\end{align}
The parameters $M_n$ and $S_n$ can be determined from the mean and variance
\begin{subequations}
\label{eqn:normalized_excitations_statistics}
\begin{align}
	\mean{\hat e_n} &= \exp\left(M_n + \frac{S_n^2}{2}\right)
\\
	\var(\hat e_n) &= e^{2M_n + S_n^2} \left(e^{S_n^2} -1\right)
	\;.
\end{align}
\end{subequations}
Solving these equations for $M_n$ and $S_n$, we obtain
\begin{align}
	M_n &= \ln\mean{\hat e_n} - \zeta
& \text{and} &&
	S_n &= \sqrt{2\zeta}
	\;,
	\label{eqn:lognorm_parameter}
\end{align}
where  $\zeta = \frac12 \ln(1 + \var(\hat e_n)\mean{\hat e_n}^{-2})$.
Eq.~{\bf 6} of the main text follows from this and Eq.~{\bf 5}.
For small $\mean{a_n}$ we have
\begin{align}
	\mean{a_n} \approx 
		\frac{2 \sqrt{\zeta/\pi } }{\ln (\alpha )+\zeta}
		\exp\left[-\frac{(\ln (\alpha )+\zeta )^2}{4 \zeta }\right]
	\;,
\end{align}
which follows from $\erfc(x) \approx e^{-x^2}/(x\sqrt\pi)$, valid for $x \gg 1$.
For small $\zeta$, we obtain the approximate scaling $\ln\mean{a_n} \sim -(\ln \alpha)^2/(4\zeta)$, where $\zeta \sim s^{-1}$ for $s \gg 1$.

\section{Odor discriminability}
We quantify the discriminability of two odors by the Hamming distance~$d$ of their respective representations~$\vect a$ for several different cases:

\paragraph{Uncorrelated odors}
The expected distance~$\mean d$ between the activity patterns~$\vect a^{(1)}$ and $\vect a^{(2)}$ of two independent odors is
\begin{align}
	\mean d &= \Nr\left(\mean{a_n^{(1)}} + \mean{a_n^{(2)}} - 2\mean{a_n^{(1)}}\mean{a_n^{(2)}}\right)
	\label{eqn:discriminability_independent}
	\;,
\end{align}
where $\mean{a_n^{(1)}}$ and $\mean{a_n^{(2)}}$ denote the expected activities of the two odors, averaged over sensitivity matrices, and we neglect correlations~$\cov(a_n, a_m)$ for simplicity.

\paragraph{Adding target to background}
We calculate the expected change~$\mean d$ of the representation when a target odor~$\vect c^{\rm t}$ is added to a background odor~$\vect c^{\rm b}$.
Because the odor concentrations are specified, we consider the actual excitations~$e_n$ instead of the normalized quantities~$\hat e_n$.
Taking an ensemble average over sensitivity matrices, the excitations associated with the two odors are characterized by probability distribution functions $f_E^{\rm t}(e^{\rm t})$ and $f_E^{\rm b}(e^{\rm b})$ for the target and the background, respectively.
We here consider log-normally distributed $e_n$, which are parameterized by their mean and variance,
\begin{align}
	\mean{e_n} &= \bar S \sumi c_i
&
	\var(e_n) &= \var(S_{ni}) \sumi c_i^2
	\;,
\end{align}
where $\var(S_{ni}) = \bar S^2(e^{\lambda^2} - 1)$.

When the target is added to the background, the expected threshold~$\mean\gamma$ increases from $\gamma^{\rm b}= \alpha \mean{e^{\rm b}}$ to $\gamma^{\rm s} = \alpha (\mean{e^{\rm b}} + \mean{e^{\rm t}})$, where $\mean{e^\kappa}$ denotes the mean excitation~$\mean{e^\kappa} = \int \! z \, f_E^\kappa(z) \, \diff z$ for $\kappa={\rm t}, {\rm b}$.
This increase in the threshold can deactivate a channel if it was previously active, \ie if its excitation was larger than the threshold associated with the background, $e^{\rm b} > \gamma^{\rm b}$.
For such $e^{\rm b}$, the probability that the receptor gets deactivated by adding the target is $P(e^{\rm b} + e^{\rm t} < \gamma^{\rm s} | e^{\rm b})$.
Integrating over all possible~$e^{\rm b}$, we thus get the probability~$p_{\rm off}$ that a channel becomes inactive,
\begin{align}
	p_{\rm off} &= \int_{\gamma^{\rm b}}^{\infty}
		\! P\left(e^{\rm b} + e^{\rm t} < \gamma^{\rm s} \,\middle | \, e^{\rm b} \right) f^{\rm b}_E\bigl(e^{\rm b}\bigr) \, \diff e^{\rm b}
\notag\\
	&= \int_{\gamma^{\rm b}}^{\infty}
	F^{\rm t}_E\bigl(\gamma^{\rm s} - e^{\rm b}\bigr)  f^{\rm b}_E\bigl(e^{\rm b}\bigr) \, \diff e^{\rm b}
	\;,
\end{align}
where $F^{\rm t}_E(e^{\rm t})$ is the cumulative distribution function associated with $f^{\rm t}_E(e^{\rm t})$.
Conversely, a channel becomes active when the additional excitation by the target odor brings it above the threshold~$\gamma^{\rm s}$.
The associated probability~$p_{\rm on}$ reads
\begin{align}
	p_{\rm on}  &= \int_0^{\gamma^{\rm b}}
	\left[1 - F^{\rm t}_E\bigl(\gamma^{\rm s} - e^{\rm b}\bigr)\right]  f^{\rm b}_E\bigl(e^{\rm b}\bigr) \, \diff e^{\rm b}
	\;.
\end{align}
Taken together, the expected number~$\mean{d}$ of channels that change their state reads
\begin{align}
	\mean d &= \Nr \cdot \bigl(p_{\rm on} + p_{\rm off} \bigr)
	\label{eqn:mixture_hamming}
	\;.
\end{align}
There are three simple limits that we can solve analytically:
If there is no target, $\mean{e^{\rm t}}=0$, the activation pattern does not change and we have $\mean d=0$.
In the opposing limit of a dominant target, $\mean{e^{\rm t}} \rightarrow \infty$, the activation patterns are independent and we recover the distance~$\mean{d}_{\rm max}$ for uncorrelated odors, which is given by \Eqref{eqn:discriminability_independent}.
Lastly, in the case where the target and the background are identically distributed, $\mean{e^{\rm b}} = \mean{e^{\rm t}}$ and $\var(e^{\rm b})=\var(e^{\rm t})$, we have $\mean d=\frac12 \mean d_{\rm max}$.

\paragraph{Discriminating two odors of equal size}
We consider the simple case of two odors that each contain $s$ ligands at equal concentration, sharing $s_{\rm b}$ of them, such that the expected threshold~$\mean\gamma$ is the same for both odors.
Similar to the derivation above, we here calculate the probability~$p$ that a channel is active for one odor, but not for the other.
The $s_{\rm b}$ ligands that are present in both odors cause a baseline excitation~$e^{\rm b}$, which is distributed according to~$f^{\rm b}_E(e^{\rm b})$.
A channel is inactive for an odor with probability $F^{\rm d}_E(\mean\gamma - e^{\rm b})$, where $F^{\rm d}_E(e^{\rm d})$ is the cumulative distribution function of the excitation caused by the $s_{\rm d}  = s - s_{\rm b}$ different ligands.
Hence,
\begin{align}
	p &= 
		2\int_0^{\mean\gamma}
			F^{\rm d}_E(z)
			\bigl[1 - F^{\rm d}_E(z)\bigr]
			f^{\rm b}_E(e^{\rm b})
		\diff e^{\rm b}
	\;,
\end{align}
where $z = \mean\gamma - e^{\rm b}$.
Note that the upper bound of the integral is~$\mean\gamma$ since channels will be active for both odors if $e^{\rm b} \ge \mean\gamma$.
The associated Hamming distance~$\mean d$ between the two odors is then given by $\mean d=p \Nr$.

\section{Receptor binding model}
We consider a simple model where receptors~$R_n$ get activated when they bind ligands~$L_i$.
This binding is described by the chemical reaction \mbox{$R_n + L_i \rightleftharpoons R_nL_i$}, where $R_nL_i$ is the receptor-ligand complex.
In equilibrium, the concentrations denoted by square brackets obey $[R_nL_i] = K_{ni} \cdot [R_n][L_i]$, where $K_{ni}$ is the binding constant of the reaction.
Hence,
\begin{align}
	[R_nL_i] &= \frac{c^{\rm rec}_n K_{ni} c_i}{1 + \sum_i K_{ni} c_i}
	\label{eqn:binding}
	\;,
\end{align}
where we consider the case where multiple ligands compete for the same receptor.
Here, $c_i = [L_i]$ is the concentration of free ligands and $c^{\rm rec}_n=[R_n] + \sum_i [R_nL_i]$ denotes the fixed concentration of receptors, which is related to the copy number of receptors of type~$n$.
We consider a simple receptor model where the excitation is proportional to the concentration of the bound ligands, such that the excitation accumulated in glomerulus~$n$ reads
\begin{align}
	e_n &= \beta_n \frac{N^{\rm rec}_n}{c^{\rm rec}_n} \sumi [R_nL_i]
		= \beta_n N^{\rm rec}_n \frac{\sum_i K_{ni} c_i}{1 + \sum_i K_{ni} c_i}
	\label{eqn:receptor_occupancy}
	\;.
\end{align}
Here, $N^{\rm rec}_n$ is the copy number of receptors of type~$n$ and $\beta_n$ characterizes their excitability, which could for instance be modified by point mutations~\cite{Yu2015}.
Defining~$S_{ni} = \beta_n N^{\rm rec}_n K_{ni}$, we recover Eq.~{\bf 1} of the main text in the limit of small concentrations, $\sum_i K_{ni} c_i \ll 1$.
The sensitivities are thus proportional to the copy number~$N^{\rm rec}_n$ and the biochemical details encoded in $\beta_nK_{ni}$.

\renewcommand{\addcontentsline}[3]{} %

\bibliographystyle{aipauth4-1}
\bibliography{bibdesk}